# Thermoacoustic Instability Suppression and Heat-Release Forcing of a Laminar Flame Using Ionic Wind

Dustin L. Cruise*, Aman Satija, Galen B. King
*Department of Mechanical Engineering, Purdue University*
*585 Purdue Mall, West Lafayette, IN, 47907, USA*

**Advancements in combustion technologies are often impeded by complex combustion dynamics. Active control has proven effective at mitigating these dynamics in the lab, but mass adoption requires more affordable, lightweight, and reliable actuators. Here, a new actuator concept is presented which utilizes sub-breakdown electric fields, the inherent plasma nature of flames, and the electrohydrodynamic effect to create flame stabilization points. These electrically controlled stabilization points allow variable distortion of a laminar flame and bidirectional forcing of the flame heat-release. The electric field-based actuator is combined with a simple feedback controller to demonstrate suppression of a thermoacoustic instability. The instability sound pressure level was reduced by 27 dB and in less than 60 ms upon enabling the controller. The use of a sub-breakdown electric field requires a mere 40 mW to stabilize a 3.4 kW thermal power flame. The absence of any moving parts and low electrical power required make this a promising actuator concept for many combustion applications.**

## 1. Introduction

Combustion, the process of burning fuels to release energy, is society's primary energy source but also stands as the largest contributor of greenhouse gas emissions. Many efforts are underway to reduce these emissions, but approaches are commonly impeded by complex combustion dynamics that manifest as difficulties with flame anchoring, flashback, or thermoacoustic instabilities (*1*). Active control, which forces the flame with various forms of actuators, has proven effective at addressing these issues but has not seen widespread adoption due to actuator weight, cost, and reliability concerns (*2*). A promising alternative utilizes electric fields to affect flames through their inherent plasma nature (*3*). Flames are weak plasmas containing charged particle densities on the order of $10^{9\text{-}10}$ cm$^{-3}$ (*4*). In CH$_4$/air flames, the dominate positive charge carrier (cation) is hydronium (H$_3$O$^+$), while the dominant negative charge carrier is the electron (*5,6*). Application of an external electric field accelerates the charged particles through the Lorentz force. Elastic collisions between the accelerated cations and neutral gas molecules result in a considerable amount of momentum transfer that alters the flow field; an electrohydrodynamic effect referred to as Ionic Wind (*7*).

In a pair of studies by Ren et al. (*8,9*), particle image velocimetry (PIV) and electric-field-induced second-harmonic generation (ESHG) captured ionic wind creating a local and controllable flow velocity reduction. Reducing the local flow velocity to the laminar flame speed caused the flame to propagate and stabilize in the induced low velocity region, similar to flame stabilization in the recirculation region of an aerodynamic bluff-body. The speed with which the flame front propagated upstream in response to a step increase of the electric field strength was also captured. Analysis showed the flame front propagated following standard laminar flame mechanics, moving at a speed equal to the difference between the laminar flame speed and the local flow velocity. This indicated that the primary effect of sub-breakdown electric fields in premixed CH$_4$/air flames is the electrohydrodynamic effect.

Corresponding author, dustin.cruise@gmail.com

The velocity reduction by ionic wind can be quite large relative to typical laminar flame speeds, with a theoretical value of 5.5 m/s (*7*) and the highest observed experimental value in $CH_4$/air flames of 1.6 m/s (*10*). This suggests that flow field perturbations created by ionic wind should see significant changes of the flame front. The same studies by Ren et al. (*8, 9*) observed a rapid change of a relatively flat profile flame to a steep inverted conical, which was attributed to the change in the flow field by ionic wind. Marcum and Ganguly (*11*) captured one of the largest distortions where a laminar conical geometry rapidly collapsed to a wrinkled laminar flamelet geometry with the application of a step in electric field magnitude (~600 V/cm). Additionally, the collapse occurred in less than 20 ms suggesting the effect could be sufficiently fast to counteract thermoacoustic instabilities with oscillation frequencies up to a few hundred Hertz. Many of the instabilities observed in practice occur in this frequency range (*2*). Modifying the flame heat-release is considered a primary means for suppressing thermoacoustic instabilities. Alternating electric fields have been demonstrated forcing large heat-release amplitudes over a wide frequency range (4 to 450 Hz) (*12*). Studying the direct application of electric fields to thermoacoustically unstable flames yields valuable insights into the coupled acoustic, combustion, and plasma dynamics. Henderson and Xu (*13*) used a Rijke tube configuration to create an instability of a premixed propane-air flame outfitted with electrodes. Applying a DC electric field was shown to increase the effective acoustic damping ratio and lower the instability amplitude. The studies to date have shown promising individual effects of electric fields on flames. However, the detailed mechanisms relating the ionic wind effect and modification of the flame heat-release or suppressing an instability have not been reported.

Here, we present an ionic wind-based mechanism for creating large structural distortions in a $CH_4$/air laminar flame. The distortions are used to bidirectionally force the flame heat-release. Coupling this heat-release forcing ability with a feedback controller allows suppression of thermoacoustic instabilities. A key feature is the use of thin cathode elements to localize the ionic wind effect and consequently allow control over the location of the ionic wind velocity reduction. By locating the velocity reduction in an initially high-velocity region, a new flame stabilization region and flame root (*2*) are created. Creation of a new flame root causes a significant distortion of the laminar flame shape. A heat-release change accompanies the distortion due to the change of the volume enclosed by the flame surface. This method is improved by using multiple thin cathodes elements that create a proportional number of distortions over the electric field range, linearizing the relationship between the applied electric field and heat-release. Benefits of this method include having no moving parts and requiring very little electrical power, requiring only 40 mW to stabilize a 3.4 kW thermal power flame. This represents a significant advancement in the manipulation of flames through their inherent plasma nature and will assist efforts to further reduce combustion emissions.

## 2. Results

### 2.1 Electrohydrodynamic Bluff-Body

An effect demonstrated for the first time in this work is the creation of what we hereafter refer to as an electrohydrodynamic (EHD) bluff-body. An EHD bluff-body creates a change in the flame that is visually identical to a classic aerodynamic bluff-body. An example of an aerodynamic bluff-body effect is shown in Fig. 1. A ring-stabilized jet burner (*14*) using a premix of methane and air created a laminar conical flame (Fig. 1A, Fig. S1). The bulk gas velocity was 1.3 m/s and the equivalence ratio was 0.95. As shown in Fig. 1B, placing a small metal rod across the exit plane of the burner acted as an aerodynamic bluff-body and caused a "V" to form in the flame surface. As depicted in Fig. 1C, the aerodynamic bluff-body causes flow detachment and recirculation (*15*) to reduce the local flow velocity to the laminar flame speed, $S_u$.

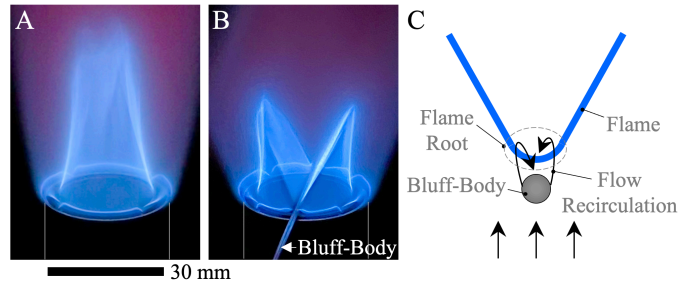

**Fig. 1. The effect of an aerodynamic bluff-body on the flame shape.** (**A**) The initial conical flame shape. (**B**) Placing a Ø1.6 mm metal rod at the base of the burner acted as an aerodynamic bluff-body and created a "V" in the flame shape. (**C**) The aerodynamic bluff-body causes flow detachment and recirculation, creating a low velocity region where a new flame root forms.

Creating an EHD bluff-body began with adding electrodes to the burner, as shown in Fig. 2A. Compared with previous studies (*10*), the key feature to increasing the sensitivity of the flame to an electric field was the attachment of a thin conductive wire across the burner face. The wire was tinned-copper electrical wire and attached by solder to the burner. An important feature was that the wire diameter was small enough (0.127 mm) so that neither flow detachment nor recirculation (*16*) occurred at the velocities tested. Thus, the wire did not act as an aerodynamic bluff-body altering the flame shape. The wire and burner were electrically grounded and served as the negative electrode (cathode). The positive electrode (anode) was constructed from a 1.6 mm diameter stainless steel rod and was located just far enough downstream to not interfere with the conical primary reaction zone; 43 mm from the burner exit plane.

Applying a DC voltage between the electrodes of sufficient magnitude (~500V) caused the flame to transition to the "V"-shape shown in Fig. 2B. This EHD "V"-flame is visually identical to the aerodynamic bluff-body example shown in Fig. 1B. A schematic of the EHD bluff-body process and flame shape change is shown in Fig. 2C. The applied electric field accelerates the hydronium ($H_3O^+$) cations towards the cathode and the electrons towards the anode. Elastic collisions between the accelerated cations and the neutral gas molecules transfer momentum to the neutral molecules, altering the velocity of the flow (i.e., ionic wind) (*7*). The specific electrode layout and polarity used created an electric field that is largely aligned against the bulk gas flow direction. As the cation current will be mostly aligned with the electric field, the cation current and ionic wind will also be against the bulk flow direction and thus will locally reduce the flow velocity. A similar burner and electrode arrangement in Altendorfner et al. (*10*) captured this local velocity reduction from ionic wind. The convergence of the electric field at the wire cathode focuses the cations toward the wire and localizes the velocity reduction around it. If the magnitude of the electric field is made large enough such that the ionic wind reduces the local flow velocity to the laminar flame speed ($S_u$), a new flame root forms creating the "V" of the flame surface.

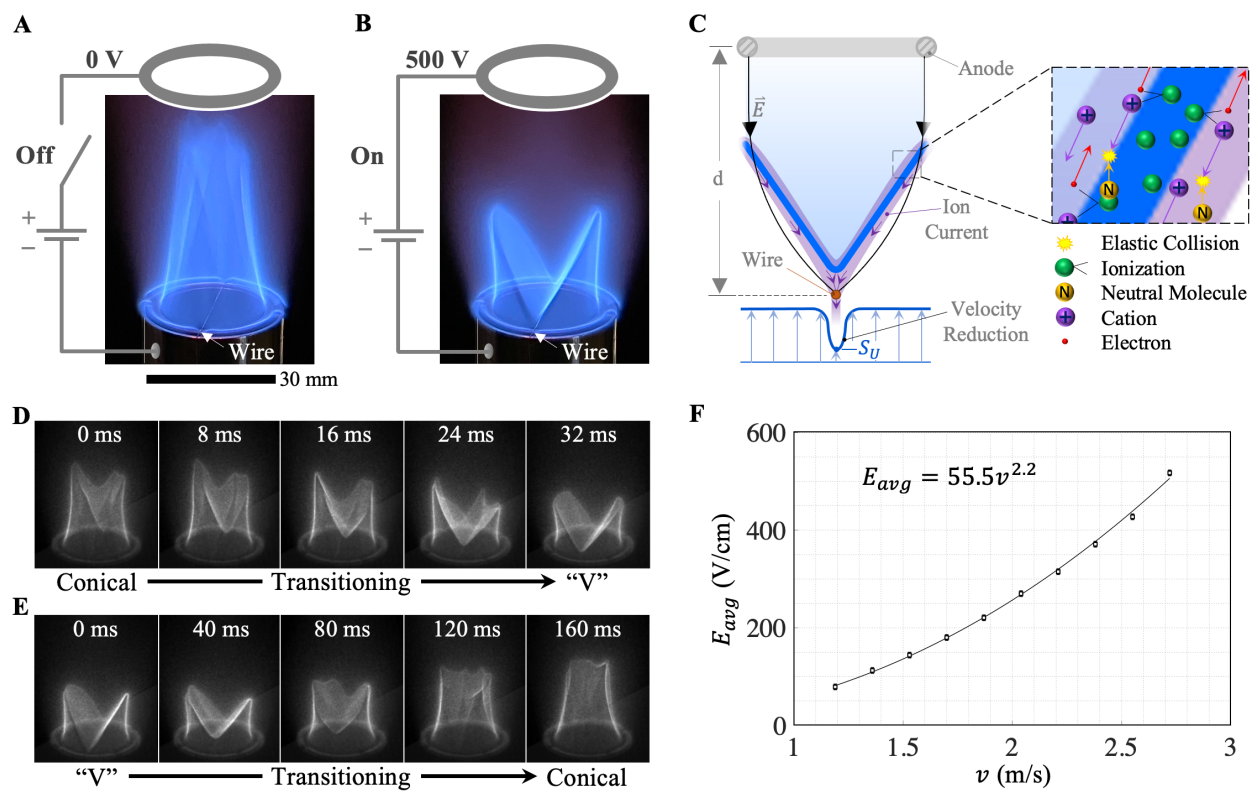

**Fig. 2. Creating an EHD bluff-body to distort the flame shape.** (**A**) A thin tinned-copper wire was attached across the burner face. With no voltage applied to the anode and thus no electric field present, the conical flame shape was unaffected by the presence of the small wire. (**B**) Applying a voltage to the anode created an electric field and caused a "V"-surface in the flame shape, which was visually identical to the previous aerodynamic bluff-body example. (**C**) Schematic of creating the EHD bluff-body: Applying a voltage to the anode creates an electric field, $\vec{E}$, that converges to the wire cathode. The ion currents follow the electric field lines and are thus focused toward the wire. The ion currents locally reduce the flow velocity through ionic wind, creating a local velocity reduction near the wire. Reducing to the laminar flame speed ($S_U$) creates a new flame root and the observed "V" in the flame structure. (**D**) High-speed images of the transition from the conical to the "V"-flame shape. (**E**) High-speed images of the transition from the "V"-shape back to the conical shape. (**F**) The minimum electric field strength required to maintain the flame in the "V"-shape vs. the bulk flow velocity.

The dynamic transition between the conical and "V"-flames was captured with a high-speed camera at 2000 frame per second (fps). The conical to "V"-flame transition is shown in the image sequence of Fig. 2D, and the reverse transition from the "V" back to the conical shown in Fig. 2E. In the forward transition of Fig. 2D, the first sign of the transition was an indentation on the flame surface near the back end of the wire. The indentation then propagated along the wire (normal to the bulk flow direction) and formed the "V"-surface as it progressed. The total transition time between shapes was around 30 milliseconds (ms). This behavior is consistent with the proposed mechanism that the electric field creates a local velocity reduction immediately downstream of the wire, into which the flame then propagates. The reverse transition from the "V"-shape back to the conical shape resembled a flame surface being convected downstream with the flow. This transition was initiated when the electric field decreased below some threshold value and took approximately 160 ms to complete. The large difference between the forward and reverse transition times is due to the forward transition being forced by the electric field and ionic wind, whereas the reverse transition is unforced and the flame front is convected with the flow back to the original conical shape.

The minimum electric field strength required to maintain the "V"-shape versus the bulk velocity is shown in Fig. 2F. The average value of the electric field, $E_{avg}$, is used here and defined as $E_{avg} = U/d$, where $U$ is the voltage difference across the electrodes and $d$ is the electrode spacing (~ 43mm). The trend in Fig. 2F shows the minimum required electric field strength scales with the velocity to the power of 2.2. Ren et. al (*8*) observed a similar quadratic relationship but between the electric current and the bulk velocity.

## 2.2 Heat-Release Change

The benefit of distorting the flame surface with the electric field is that it causes a brief change in the flame heat-release. As depicted in Fig. 3A, when the flame surface transitions from the conical to the "V"-shape, the internal volume enclosed by the flame surface is reduced. The volume difference between the shapes is consumed during the flame shape transition and the chemical potential energy of the volume difference adds to the flame heat-release. A control volume analysis is used to model this effect. As shown in Fig. 3B, the control volume is the internal volume enclosed by the flame surface and the exit plane of the burner tube. Mass enters the control volume through the exit plane of the burner tube at a constant rate of $\dot{m}_{in} = \rho_m A_b v$, where $\rho_m$ is the mixture density, $A_b$ is the burner tube exit area, and $v$ is the bulk velocity. Mass leaves the control volume through the flame surface at a rate of $\dot{m}_{out}(t) = \rho_m A_f(t) S_u$, where $A_f(t)$ is the flame surface area and $S_u$ is the laminar flame speed. Applying mass conservation to the control volume gives

$$\dot{m}_{out}(t) = \rho_m A_b v - \rho_m \frac{dV(t)}{dt} \tag{1}$$

where the second term on the right-hand side is the rate of change of the control volume mass. The flame heat-release is related to the mass flow rate out of the control volume (through the flame surface) by the heat of combustion: $q(t) = h_c \dot{m}_{out}(t)$ (*15*). Applying this to Eq. 1 gives the heat-release in terms of the control volume change

$$q(t) = h_c \rho_m A_b v - h_c \rho_m \frac{dV(t)}{dt} \tag{2}$$

Suppressing a thermoacoustic instability by actively forcing the heat-release requires creating a fluctuating component of heat-release (*2*), defined as

$$q'(t) = q(t) - \bar{q} \tag{3}$$

where $\bar{q}$ is the time-average heat-release. Combining Eq. 2 and 3 gives the fluctuating heat-release in terms of the changing control volume

$$q'(t) = - h_c \rho_m \frac{dV(t)}{dt} \tag{4}$$

This shows that changing the control volume by the electric field will modify the heat-release. One perspective is that the volume enclosed by the flame surface is an energy reservoir and by moving the flame surface with the electric field, the energy reservoir can be expended or replenished to briefly change the flame heat-release.

The flame heat-release was measured to capture its change during the transition between flame shapes. CH* chemiluminescence is considered a good marker of heat-release (*17-19*) and was measured using a photomultiplier tube (PMT) with a 10 nm bandpass filter centered at 430 nm.

The PMT was located far enough from the flame so that the entire flame was in the field of view of the PMT. When the flame and its heat-release are fluctuating, the global heat-release variation is proportional to the CH* intensity variation (*12, 20*):

$$\frac{I(t) - \bar{I}}{\bar{I}} = \frac{q(t) - \bar{q}}{\bar{q}} \qquad (5)$$

where $I(t)$ is the measured CH* intensity and $\bar{I}$ is the time-average intensity.

A triangular electric field waveform was applied to the flame, with the single-wire cathode, to cause flame shape transitions and accompanying heat-release events, as shown in Fig. 3C. The heat-release is plotted as the percent deviation from the mean value, $q'/\bar{q}$ (%). Application of the triangular electric field waveform caused a repeatable change in heat-release, with a single positive pulse when the electric field was increasing and a single negative pulse when it was decreasing. The single positive pulse of heat-release was around 14-16% in amplitude and occurred when the increasing electric field triggered a flame shape transition from the conical to the "V"-shape. The connection between the positive heat-release pulse and the flame shape transition was confirmed by comparing the heat-release plot with the synchronized images of the flame shape transition, shown in Fig. 3D. The heat-release changed only when the flame shape was in transition (Fig. 3D, ii, iii, and iv). When the flame shape was stationary (Fig. 3C, i and v), there was little to no change in the heat-release. These observations agree with the relationship of Eq. 4 which showed that only a changing control volume would produce a heat-release change. Similarly, a single negative pulse around 5 – 7% in amplitude occurred when a decreasing electric field triggered a flame shape transition from the "V" back to the conical shape. Comparison of the negative pulse with its corresponding images (Fig. 3E) showed the same relationship expected by Eq. 4. The pulse was negative because transitioning from the "V" back to the conical shape resulted in the control volume increasing ($dV/dt > 0$). Video of the flame transitions and heat-response to a square electric field waveform is provided in the Supplementary Material.

An alternative way to characterize the relationship between electric field and heat-release is to integrate the heat-release signal. Using Eq. 4 and 5, it can be shown that integrating the deviation of the CH* chemiluminescence intensity, $I'(t)$, gives a value proportional to the energy expelled from the control volume that modifies the heat-release

$$\int_0^t \frac{I'(\tau)}{\bar{I}} d\tau = \int_0^t \frac{q'(\tau)}{\bar{q}} d\tau = \frac{-h_c \rho_m \Delta V(t)}{\bar{q}} = \frac{1}{\bar{q}} \Delta Q(t) \qquad (6)$$

We refer to the quantify $\Delta Q(t)$ as the *thermal energy change*. This quantity is plotted against the electric field strength in Fig. 3F and both axes are normalized as only the shape of the profile is of interest. The profile is a bistable relationship with hysteresis between the forward and reverse directions. Starting from zero and increasing the electric field strength, the thermal energy change ($\Delta Q$) is relatively constant until the flame transitions from the conical to the "V"-shape (indicated on the plot). The positive heat-release pulse during the shape transition creates this step increase in the thermal energy change. Increasing the electric field further sees little change of the thermal energy, all the way to the maximum electric field value. Upon decreasing the electric field, there is little change in the profile until the flame transitions from the "V" back to the conical shape. Here, the negative pulse of heat-release creates a step decrease in the thermal energy change.

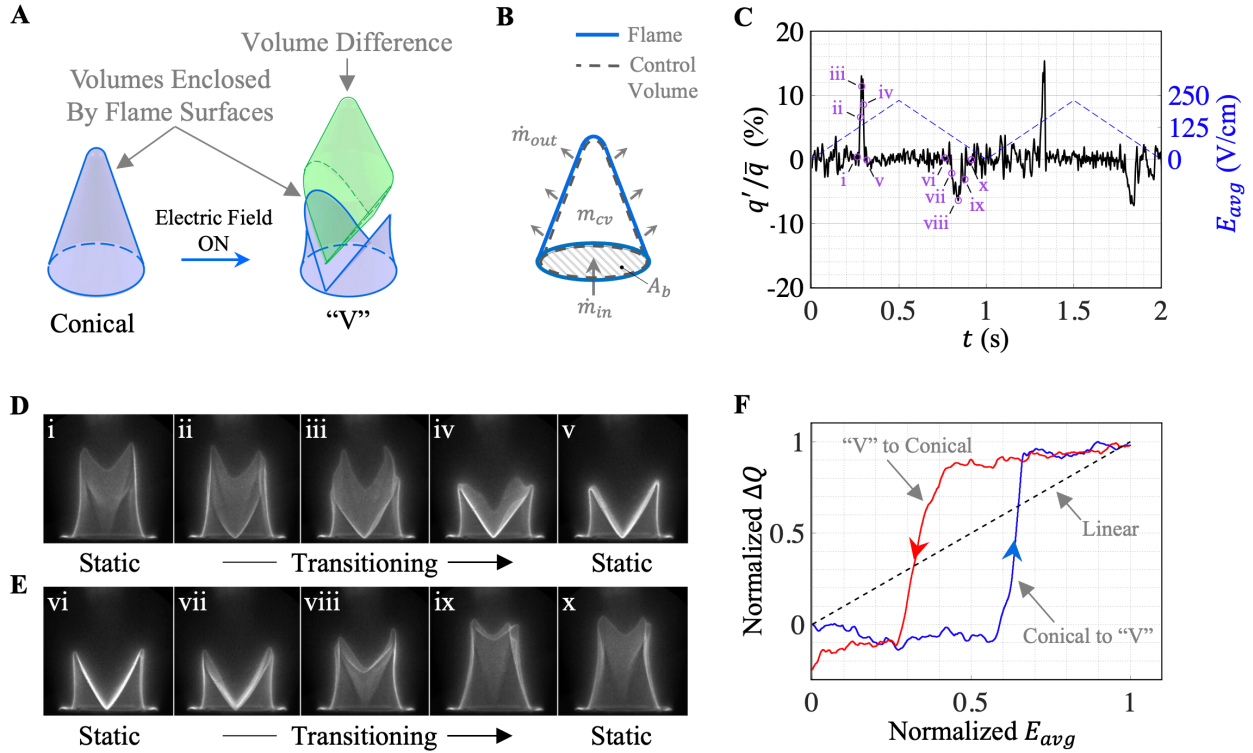

**Fig. 3. Heat-release change during flame shape transition.** (**A**) Volumes enclosed by the conical and "V"-flame shapes, which are different; the volume difference is consumed during the transition. (**B**) Control volume defined by the flame surface and the burner exit plane ($A_b$). Mass enters the control volume, $\dot{m}_{in}$, leaves the control volume, $\dot{m}_{out}$, and the control volume contains a quantity of mass, $m_{cv}$. (**C**) Measured change in heat-release when applying a triangular electric field waveform to trigger flame shape transitions. (**D**) Images of the conical to "V"-flame transition. These are related to the points in (C). (**E**) Images of the "V" to the conical transition. These are also related to the points in (C). (**F**) Normalized electric field magnitude vs. thermal energy change.

## 2.3 Multi-Element Cathode

The heat-release response and thermal energy profile of Fig. 3, C and F, demonstrated that the electric field could affect the flame heat-release. However, there was only a single change over the entire electric field range and there was significant hysteresis between the forward and reverse directions. The ideal relationship between the electric field and the thermal energy change would be linear, where incremental changes in the electric field would cause proportional and incremental changes in the thermal energy ($\Delta Q$). This isn't directly possible due to the bistable nature of the EHD bluff-body formation process and the resulting abrupt transitions of the flame shape. However, it was found that the desired linear relationship can be approximated by adding more cathode elements to cause a proportional number of EHD bluff-bodies and corresponding flame shape transitions. Each flame shape transition causes a step in the thermal energy change, as conceptualized in Fig. 4A. Therefore, adding $N$ individual cathode elements will create $N$ flame shape transitions and $N$ steps in the thermal energy change vs. electric field profile of Fig. 4A.

A cathode consisting of three wires was constructed to demonstrate the concept of multiple cathode elements (Fig. 4B). As shown by the flame images in Fig. 4C, the flame experienced three shape transitions over the electric field range. Each transition was due to an EHD bluff-body forming over one of the three wires ("front view" of Fig. 4C). A triangular electric field waveform was applied to induce shape transitions and a heat-release variation, as shown in Fig. 4D. The response showed three positive heat-release pulses (t1,2,3) when the electric field was increasing and three negative pulses (t4,5,6) when the field was decreasing. Each pulse occurred when the flame shape was transitioning, due to the formation or dissipation of an EHD bluff-body over a wire. Video of the flame transitions and heat-response to a triangular electric field waveform is provided in the Supplementary Material. The heat-release pulses created steps in the thermal energy change profile, shown in Fig. 4E. The three positive heat-release pulses (t1,2,3) created three positive steps in the profile, while the three negative pulses (t4,5,6) created three negative steps. This three-wire cathode demonstrates that adding cathode elements (i.e. wires) causes a proportional number of EHD bluff-bodies and flame transitions to occur. This is beneficial for the goal of making an actuator because the same electric field range causes more heat-release events, and the control volume is reduced in steps. The $\Delta Q$ vs. $E_{avg}$ profile (Fig. 4E) reflects the multiple events and better approximates the desired linear relationship.

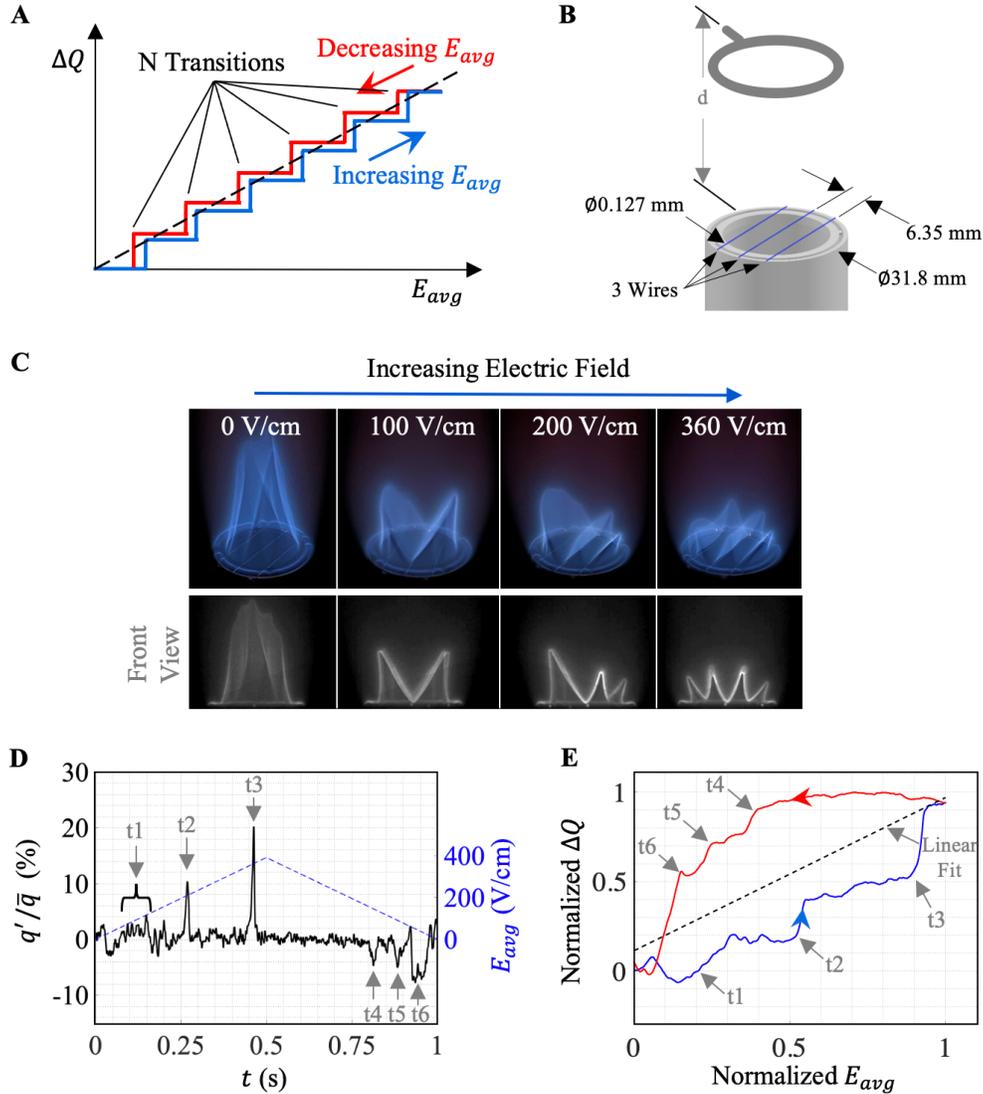

**Fig. 4. Flame transitions with a three-element cathode.** (**A**) Conceptually, each additional cathode element creates an additional flame shape transition and step in the $\Delta Q$ vs. $E_{avg}$ profile. Multiple transitions approximate the desired linear relationship. (**B**) Burner with three small diameter wires acting as the cathode elements. (**C**) The progression of flame shapes with increasing electric field. The "V" surfaces formed directly over the cathode wires. (**D**) The measured heat-release response to a triangular electric field waveform had three positive heat-release pulses while increasing the electric field magnitude (t1,2,3), and three negative pulses while decreasing the electric field magnitude (t4,5,6). (**E**) Normalized electric field magnitude vs. thermal energy change showing three positive steps followed by three negative steps.

More cathode elements were added to further improve the relationship between the electric field and the thermal energy change. Doing so with wires was impractical so instead a hexagonal metal structure (i.e., honeycomb) was used, as shown in Fig. 5A. The thin walls of the hexagonal cells act as the cathode elements where the EHD bluff-bodies form. To facilitate comparison with the previous ring-stabilized jet burners, the same air flow rate and equivalence ratio were used. The change in flame shape with increasing electric field strength is shown by the images in Fig. **5**C. The initial flame shape before applying the electric field was conical, similar in size and shape to that of the ring-stabilized jet burner. As the electric field was increased from zero, the first visible change was fine-scale wrinkling around the base of the flame (image "79 V/cm"). The wrinkling was caused by EHD bluff-bodies forming over the honeycomb cell walls. The shape of the flame was stable and didn't visibly change as long as the electric field magnitude was held constant. As the electric field was increased, the number of EHD bluff-bodies and surface wrinkling increased, as evident by the progression seen between the "79 V/cm", "158 V/cm", and "236 V/cm" images. These small and incremental changes between successive flames shapes was closer to the ideal behavior because the control volume was incrementally reduced over many steps, as opposed to the large distortions seen with the single- and three-wire cathodes. Higher electric field strengths eventually caused large distortions of the flame shape, such as those seen in the "236 V/cm", "314 V/cm", and "394 V/cm" images. Both the fine-scale wrinkling and large-scale distortions exhibited the characteristic "V" structure near the honeycomb surface. Viewing the flame from the front perspective (Fig. 5B) shows the "V" structures.

The many incremental changes of the flame shape over the electric field range changed the heat-release response from a few discrete pulses to regions of largely steady output, as shown in Fig. 5D. For example, the heat-release had a relatively steady output from 0.05 to 0.30 seconds when increasing the electric field. This region of steady positive heat-release created a quasilinear region in the thermal energy change profile of Fig. 5E, marked by the points i – iii. Similarly, there was a region of relatively steady negative heat-release (points iv – vi) when the electric field was decreasing. There were some large pulses of heat-release at the higher electric field magnitudes, such as the positive pulses in the region of 0.30 to 0.50 seconds, and the negative pulse around 0.875 seconds. These occurred when the flame experienced large distortions of the shape instead of fine-scale wrinkling. However, the profile of Fig. 5E has a quasilinear response for low to mid-range electric field magnitudes and this is ideal for creating an electric field-based actuator of heat-release. Video of the flame transitions and heat-response to a square electric field waveform is provided in the Supplementary Material.

The three cathodes tested demonstrated: 1) the ability to create an EHD bluff-body, 2) distorting the flame shape caused a heat-release change, and 3) adding cathode elements created more EHD bluff-bodies over the electric field range. When the number was significantly increased with the many surfaces provided by the honeycomb cathode, an ideal relationship between the applied electric field and resulting heat-release was approached.

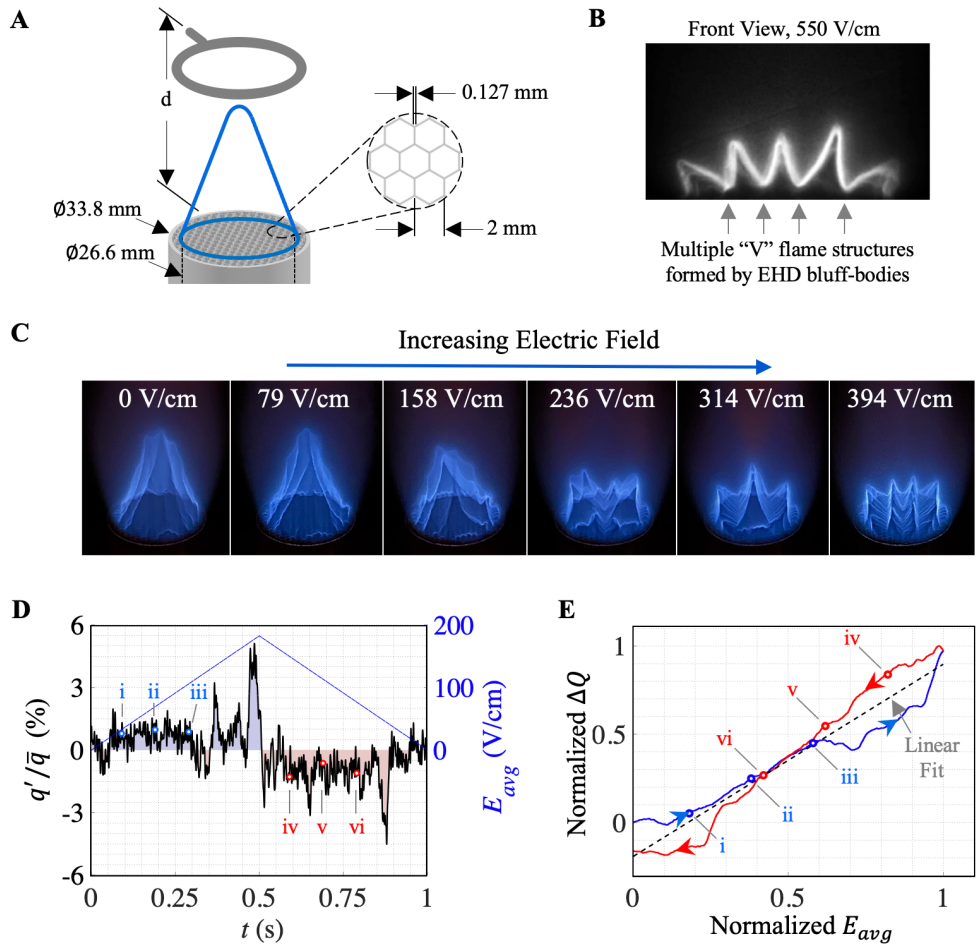

**Fig. 5. Flame transitions with a honeycomb cathode.** (**A**) Burner with a metal honeycomb acting as the cathode element. (**B**) Front view of the flame shows the heavily distorted shape at high electric field magnitude was a series of "V" flames created by multiple EHD bluff-bodies. (**C**) The progression of flame shapes with increasing electric field. (**D**) The measured heat-release response to a triangular electric field waveform. (**E**) Normalized electric field magnitude vs. thermal energy change.

*2.4 Dynamic Response*

An impulse response of the electric field on the flame and heat-release was performed to characterize the dynamic relationship between the two. The impulse was performed by applying a 1 ms wide, 330 V/cm electric field pulse, as shown in Fig 6, A and B. The flame images in Fig 6B show the creation of a local inward distortion of the flame surface by the electric field, first appearing around 5 ms after the pulse was applied. The inward direction of the distortion is consistent with the behavior seen in the honeycomb flame images (Fig. 5C), which showed that the increasing electric field made the flame more compact (through the creation of EHD bluff-bodies), and caused the flame surface to propagate inwards. Even though the pulse was only 1 ms in duration, the surface distortion it created lasted much longer, taking approximately 20 ms to travel the entire length of the flame. Tracking the axial location of the distortion (yellow arrows) in each frame showed that it traveled at approximately the bulk flow velocity of 1.3 m/s. This suggests that after the distortion was created by the electric field pulse, it was convected by the bulk flow until it reached the flame tip. This same convection of surface distortions also occurs in those created by acoustic velocity forcing (*20 – 24*).

The heat-release response to the electric field pulse is shown in Fig 6A. To reduce the impact of inherent heat-release fluctuations and better capture the variation only due to the applied pulse, an ensemble average of 8 responses was used. Comparing the heat-release response with the flame images showed that the heat-release was affected the entire time the distortion was present on the flame surface. Immediately after the pulse was applied, the heat-release increased to a steady value around 1.5% and lasted for almost 20 ms. The increased heat-release indicated the distortion caused a steady decrease of the control volume as it propagated. Around t = 20 ms, the distortion reached the tip of the flame and caused a reactant pocket to break from the flame surface. This corresponded to the end of the positive heat-release region. The final images from 21 to 27 ms show the main flame surface expanding back to the original conical shape. The expanding flame shape causes the control volume to increase and the heat-release to decrease, as seen in the heat-release response between 20 to 35 ms. This behavior of the distorting flame surface and heat-release is consistent with the proposed control volume relationship (Eq. 4).

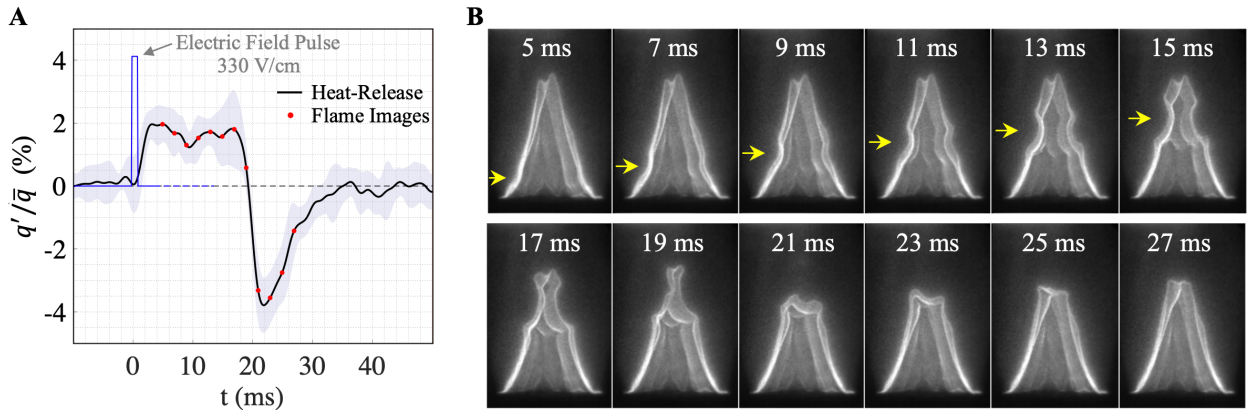

**Fig 6. Heat-release and flame response to electric field pulse.** (**A**) Ensemble mean heat-release response composed from 8 separate responses. The shaded regions represent one standard deviation. (**B**) Sequential images showing the downstream convection of the surface distortion (yellow arrows) created by the electric field pulse.

Forcing the flame with a sinusoidal electric field created a largely sinusoidal heat-release response at the same frequency, as shown in Fig. 7A. The amplitude of the sinusoidal heat-release varied significantly with the electric field forcing frequency, due to the underlying dynamic relationship between the two. A magnitude response was performed to characterize this dynamic relationship by applying a sinusoidal electric field of the form

$$E_{avg}(t) = A\sin(2\pi f_i t) + E_{DC} \qquad (7)$$

at discrete frequencies over the range of 3 to 299 Hz, in 2 Hz increments. To preserve dynamic similarity between the original conical flame and the electric field forced flame, the DC offset value was limited to 110 V/cm so the flame was only slightly distorted from the conical shape (similar to the first two images of Fig. 5C). The sinusoidal amplitude was set equal to the DC offset so that the applied electric field, $E_{avg}(t)$, was always positive.

The measured magnitude response of Fig. 7B shares similarities with the previous electric field study of Lacoste et al. (*12*) and acoustic forcing studies (*20 – 24*). Most importantly, the response showed the ability to force significant heat-release amplitudes, with a peak of 7% at 29 Hz. The magnitude response did show a strong frequency dependence. The initial magnitude from 3 to 15 Hz rises at a relatively constant rate of one order of magnitude per decade, characteristic of a derivative relationship (*25*). This behavior can be shown to be consistent with a linear relationship between the applied electric field and thermal energy change. The ideal linear relationship between the two can be modelled as

$$\Delta Q(t) = \alpha E_{avg}(t) \tag{8}$$

where $\alpha$ is an unknown proportionality constant. Given that the heat-release is the derivative of the thermal energy change (Eq. 6), the ideal relationship between a sinusoidal electric field (Eq. 7) and the heat-release is

$$q'(t) = \alpha \frac{d(E_{avg}(t))}{dt} \tag{9}$$

and

$$q'(t) = 2\pi \alpha A f \cos(2\pi f t) \tag{10}$$

Equation 9 shows a derivative relationship between the electric field and heat-release. Equation 10 shows that for a sinusoidal electric field, the heat-release will be sinusoidal with an amplitude of $2\pi \alpha A f$. The presence of the forcing frequency ($f$) in the heat-release amplitude causes the observed rise of the amplitude in the 3 to 15 Hz region of Fig. 7B.

At frequencies greater than 15 Hz, the heat-release amplitude quickly deviated from the derivative relationship of Eq. 10 and developed into a steady roll-off and series of regularly spaced local minima. This phenomenon has been observed in acoustically forced laminar flames (*20 – 24*) and is due to the finite time it takes a disturbance to propagate along the flame surface. A consequence of the finite propagation time is that multiple distortions can simultaneously be present on the surface, such as both inward and outward distortions when sinusoidally forcing the flame (*20 – 24*). The inward and outward distortions have opposite contributions to the total surface area and heat-release, and their contributions will largely cancel each other. Perfect cancellation can occur at specific frequencies and causes the fluctuating portion of heat-release to tend towards zero, thereby creating the local minima observed in the magnitude response. As the surface distortions are convected along the flame like propagating waves, the frequencies of perfect cancellation follow basic wave-propagation mechanics. Thus, the frequencies of the local minima are integer multiples of a fundamental frequency (*20, 23*).

Multiple distortions occurred when forcing the flame with sinusoidal electric fields, as shown in the flame image in Fig 7C. Multiple inward and outward distortions can be seen when compared to the time-averaged mean profile (pink). Additionally, the local minima in the magnitude response of Fig. 7B appear to be at the integer multiples of a fundamental frequency. The first local minimum occurs at 59 Hz and the rest occur at frequencies that are close to integer multiples of the 59 Hz fundamental. The presence of these regularly spaced local minima implies that the high frequency behavior of the heat-release when forcing by the electric field is dominated by flame surface mechanics (*20, 23*). Thus, the magnitude response provides little information on the upper frequency limit imposed by the electrohydrodynamic effect, other than it must be greater than that observed in Fig. 7B. This is important as many combustion dynamics of practical interest occur at frequencies greater than the 200 – 300 Hz bandwidth limit observed in Fig. 7B. Further modeling and testing will be required to determine the frequency limit imposed by the electrohydrodynamic effect.

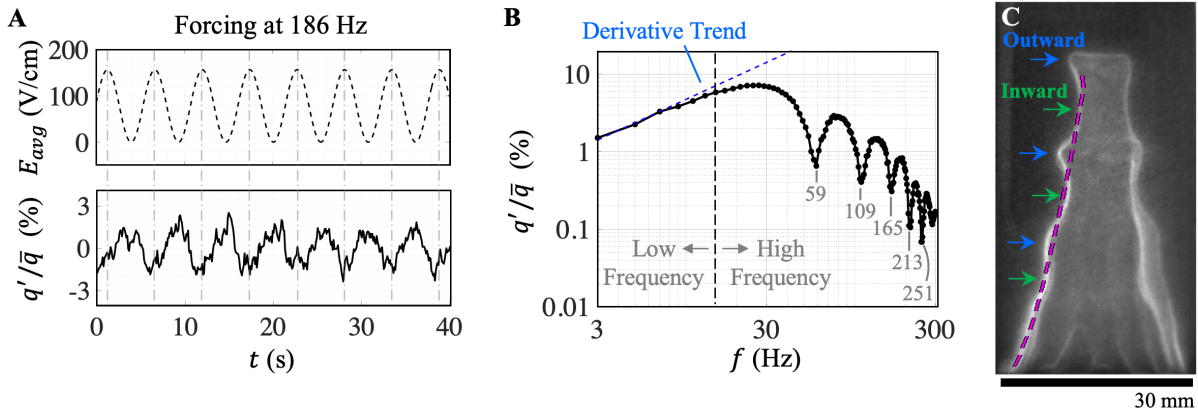

**Fig. 7. Sinusoidal forcing of the flame.** (**A**) Example of forcing the flame with a sinusoidal electric field at 186 Hz and the resulting heat-release, which was largely sinusoidal albeit with some distortion and a phase shift. (**B**) Magnitude response showing a low frequency behavior following a derivative trend, and a high frequency behavior of steady roll-off and a series of local minima. (**C**) Image of the flame being forced with the electric field at 186 Hz where multiple inward and outward surface distortions are simultaneously present. The time-averaged mean profile is included for reference (pink).

## *2.5 Suppression of a Thermoacoustic Instability*

The electric field actuator was applied to a thermoacoustic instability in order to demonstrate the efficacy of the electric field method and to explore the system level behavior. A simple thermoacoustic instability was created by enclosing the burner in a long tube, as shown in Fig. 8A. The tube acted as the acoustic chamber needed to create the standing pressure wave that couples with the flame to form the thermoacoustic instability. The specific configuration of the tube and burner used was the Rijke configuration, having open acoustic boundaries on each end and the heat source placed at the quarter wave point. An instability was found by varying the fuel and air flow rates while monitoring the acoustic pressure via a microphone. A large instability, around 134 dBA in amplitude, was found at 141Hz. The corresponding bulk flow velocity was 1.8 m/s and the equivalence ratio was 1.04. This frequency corresponded to the first longitudinal mode of the acoustic cavity. The amplitude and frequency of this instability are comparable with other laboratory-scale instability studies (*26, 27*) and was an appropriate test for the electric field actuator.

To suppress the instability with the electric field, a feedback controller was needed to synchronize the electric field to the oscillating acoustic pressure. As shown in the diagram of Fig. 8A, the acoustic pressure near the flame was measured by a microphone (GRAS 46BD) and sampled by a National Instruments myRIO data acquisition device (DAQ). The feedback controller was implemented in a software loop with a sampling frequency of 10 kHz. This sampling frequency was chosen in order to be sufficiently greater than the thermoacoustic instability frequency ($f_S > 10 f_{TAI}$). The calculated control signal was outputted by a digital-to-analog convertor (DAC). A high-voltage linear amplifier (APEX PA-99) amplified the control signal to set the anode voltage. The increased flow velocity (1.3 to 1.8 m/s) created a slightly longer flame in comparison to the flames discussed in the previous sections. Therefore, the electrode spacing was increased to 64 mm to keep the anode out of the conical primary reaction zone. The relationship between the applied anode voltage and the average electric field strength was $E_{avg}(t) = U(t)/6.4$ (V/cm).

A phase-shift controller structure was used as the feedback controller. Such controllers are often used for thermoacoustic instability suppression (*28*) because their structure is well suited for compensating unstable oscillators and they can be manually tuned while the experiment is running. The phase-shift controller structure was

$$E_{avg}(t) = Kp(t-\tau) + E_{DC} \tag{11}$$

where $K$ was the controller gain, $p(t)$ was the measured pressure, $\tau$ was the adjustable time-delay, and $E_{DC}$ was a DC offset (~79 V/cm). The controller parameters ($K$, $\tau$) were manually adjusted with the experiment running to find the optimal set that maximized the sound level reduction and gave the shortest transient settling time. The final values were $K = 6.25\ V \cdot cm^{-1} Pa^{-1}$ and $\tau = 0.6$ ms. Additional details on the system-level behavior of the closed-loop controller are given in Cruise et. al (*29*).

An example of suppressing the instability with the electric field is shown Fig. 8B. With the controller initially disabled, the flame experienced the full amplitude instability, and the acoustic pressure and heat-release were in-sync; this was expected by the Rayleigh criterion (*30*). Upon enabling the controller at t = 0 ms, the pressure oscillation immediately began diminishing and was fully suppressed in less than 60 ms. The suppression time was partially slowed by the electric field being saturated for the first 45 ms. The observed saturation and clipping of the electric field time-series stemmed from voltage limitations of the high-voltage amplifier. Increasing the upper voltage limit could further reduce the suppression time, but the 8 oscillation cycles to suppress is comparable to previous thermoacoustic actuator studies (*26*). The acoustic spectra for the instability and controlled cases are shown in Fig. 8C. The acoustic spectrum units are acoustic decibels (dBA). Comparison of the uncontrolled and the controlled spectra shows that the electric field-based actuator and feedback controller reduced the peak pressure amplitude by 27 dBA. The controlled spectrum also shows the familiar peak splitting phenomenon (*1*). More advanced control algorithms would result in further sound pressure level reductions, but they are beyond the scope of this study.

The action of the electric field was evident in how the phase relationship between the pressure and heat-release changed when the controller was enabled. As soon as the controller was enabled t = 0 ms, the synchronization between pressure and heat-release was disrupted. By t = 20 ms, the heat-release was being forced by the electric field to be 180° out-of-phase relative to the pressure. This is the theoretical fastest way to reduce the acoustic energy. This behavior is consistent with the proposed EHD mechanism which is that the electric field forces heat-release through distorting the flame shape. With a properly tuned controller, the electric field-driven component of heat-release offsets and nullifies the component of heat-release driven by the instability. Video is provided in the Supplementary Material of a similar transient controller test.

In addition to good suppression performance, the electric field method used very little electrical power to control the instability. The electrical power applied to the flame is the product of the electrode voltage and the current through the flame. Figure S2 shows the time-series of the pressure, anode voltage, current, and electrical power when suppressing the thermoacoustic instability. A peak electrical power consumption of 390 mW occurred when the controller was first enabled but quickly diminished to an average of 40 mW once the instability was suppressed.

Compared with the 3.4 kW thermal power of the flame, this electrical power is only 0.0012% of the thermal power.

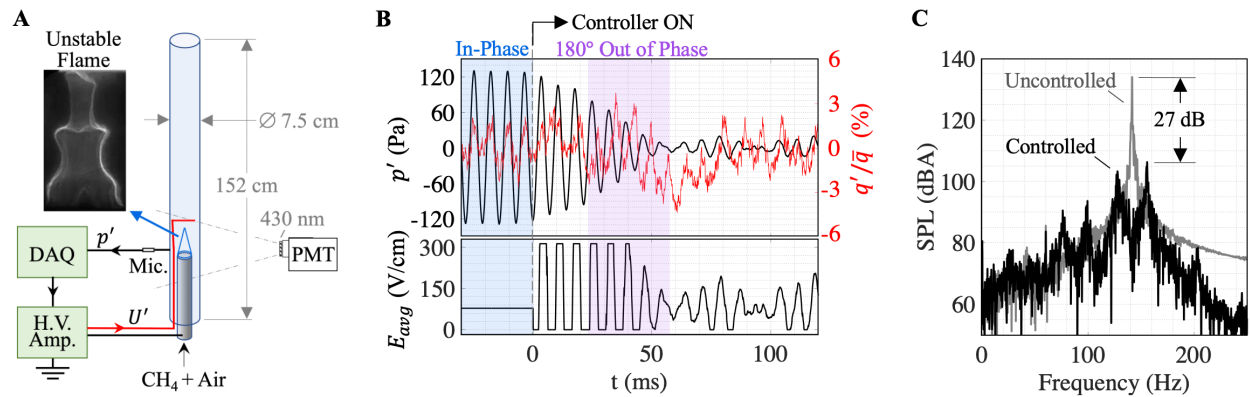

**Fig. 8. Suppression of a thermoacoustic instability with the electric field and feedback control.** (**A**) The Rijke tube setup for creating the thermoacoustic instability. (**B**) Time-series of the pressure, heat-release, and electric field showing the suppression of the instability once the controller was enabled at 0 ms. (**C**) Acoustic pressure spectra showing that the controller and electric field reduced the instability amplitude by 27 dB.

## 3. Discussion

This work has demonstrated a new mechanism for manipulating flames with electric fields. A key element is using a thin cathode geometry that converges the electric field and focuses the cation current, creating a localized velocity reduction of the flow through the ionic wind effect. Using this effect to reduce the flow velocity down to the laminar flame speed allows the formation of a new flame root and "V"-surface in the flame shape; an effect we call an EHD bluff-body for its similarity to aerodynamic bluff-body stabilization. The primary benefit for distorting the flame surface is an accompanying heat-release change. Significantly increasing the number of cathode elements via a simple honeycomb material greatly increased the number of EHD bluff-bodies and linearized the relationship between electric field and heat-release. Coupling this electric field-based actuator of heat-release with a simple feedback controller allowed suppression of a laboratory-scale thermoacoustic instability.

The performance of this method, coupled with its potentially high reliability and affordability, presents an opportunity to apply it to the most cost-sensitive combustion devices, like residential gas boilers, furnaces, and water heaters. Collectively, these devices account for 15% of U.S. annual natural gas usage (*31*) and are facing stricter emissions standards (e.g., California Rule 1111). While the use of premixed and leaner reactant mixtures is effective at reducing $NO_X$ in these devices, adverse combustion dynamics like thermoacoustic instabilities are more prevalent at these same conditions. The electric field method could stabilize these unwanted dynamics and unlock even more substantial emissions reductions.

Further development will be required to apply this method to industrial burners and gas turbine engines, characterized by higher flow velocities and turbulent flame structures. A central question is the maximum achievable velocity reduction by ionic wind, and how does that vary with pressure, fuel type, equivalence ratio, etc. The placement of the anode in the flame will not be acceptable for many applications, but detailed numerical simulation and measurement of the electric field and flow field conditions could reveal alternative electrode configurations. Given the performance and features presented here, the electric field method is a promising candidate for addressing key dynamic issues and warrants further investigation.

4. **Materials and Methods**

*Burners*

The dimensions of the ring-stabilized jet burner (*14*) used for the single and three-wire experiments are provided in Fig. S1. For the honeycomb burner of Fig. 5A, Hastelloy® honeycomb material was used to withstand the high temperatures encountered near the flame. To facilitate comparison between the flames of the ring-stabilized jet and honeycomb burners, the flow areas at the burner exit planes were made as similar as possible to one another. The flow area of the honeycomb cathode was set using an upstream flow restrictor with a 26.6 mm diameter opening. The air and methane flow rates into the burners were metered individually with Porter Model 203A mass flow controllers. The gasses were mixed far upstream of the burner and flame to ensure a full premixed state at the flame.

*High-Voltage Amplifier*

The anode voltage was controlled by an Apex PA99 high-voltage linear amplifier. Important features of the amplifier for this work are an output range of 0 to 2.5 kV, a 50 mA continuous output current, a 5 kHz power bandwidth, and maximum slew rate of $30\ V/\mu s$. The amplifier was mounted to an Apex EK36 evaluation board and the amplifier gain set by the board components to the default of 250:1. An Acopian 2.5 kV, 200 mA DC power supply was used for the amplifier positive supply rail. The amplifier negative supply rail was set to -60 V by a standard DC power supply. This negative voltage is needed to allow the amplifier output to be set to 0 Volts. The negative input to the differential amplifier was grounded while the positive input was supplied from the NI myRIO. The myRIO could not directly drive the 50 ohm input impedance of the Apex amplifier, so a TTI WA301 wideband amplifier was used as a voltage buffer between the two devices.

*Chemiluminescence Measurement and Flame Imaging*

A Hamamatsu R928 photomultiplier tube (PMT) with a 10 nm bandpass filter at 430 nm was used to measure the CH* chemiluminescence intensity. The PMT was located far enough from the flame so that the entire flame was within the PMT field of view. High-speed images were used to capture the flame transitions at camera rates of 2000 frame per second (fps). The flame was imaged using a Jenoptik 105 mm f/4.5 lens, a LaVision IRO image intensifier, and a Photron FASTCAM SA4 camera. A trigger was used to synchronize the start of camera acquisition with the data acquisition system

**Acknowledgements**
The authors thank Prof. Robert P. Lucht and Purdue Zucrow Labs for use of the high-speed camera and intensifier. The authors are grateful to Bert Gramelspacher and the ME Electronics Shop for their assistance with the electronics design of this experiment.

# Supplementary Materials for

## Thermoacoustic Instability Suppression and Heat-Release Forcing of a Laminar Flame Using Ionic Wind

Dustin L. Cruise*, Aman Satija, Galen B. King

*Corresponding author. Email: dustin.cruise@gmail.com (D.L.C.)

**Figures**

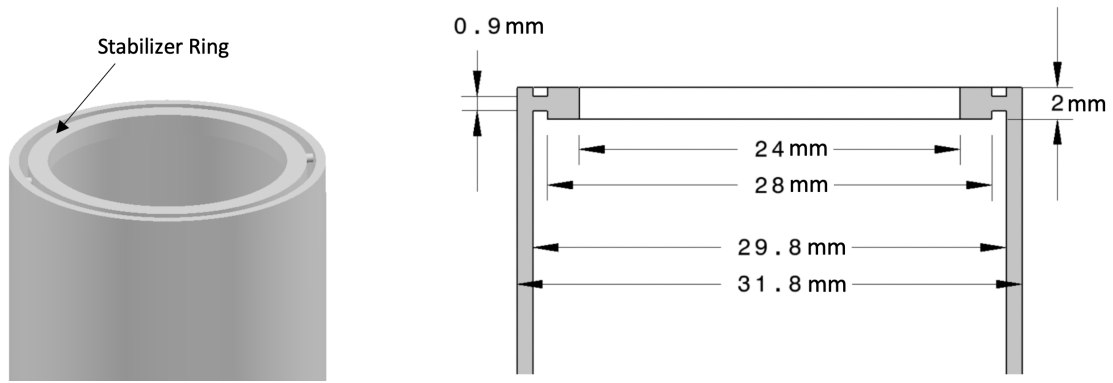

**Fig. S1. Ring-stabilized jet burner, isometric view and dimensions.**

Corresponding author, dustin.cruise@gmail.com

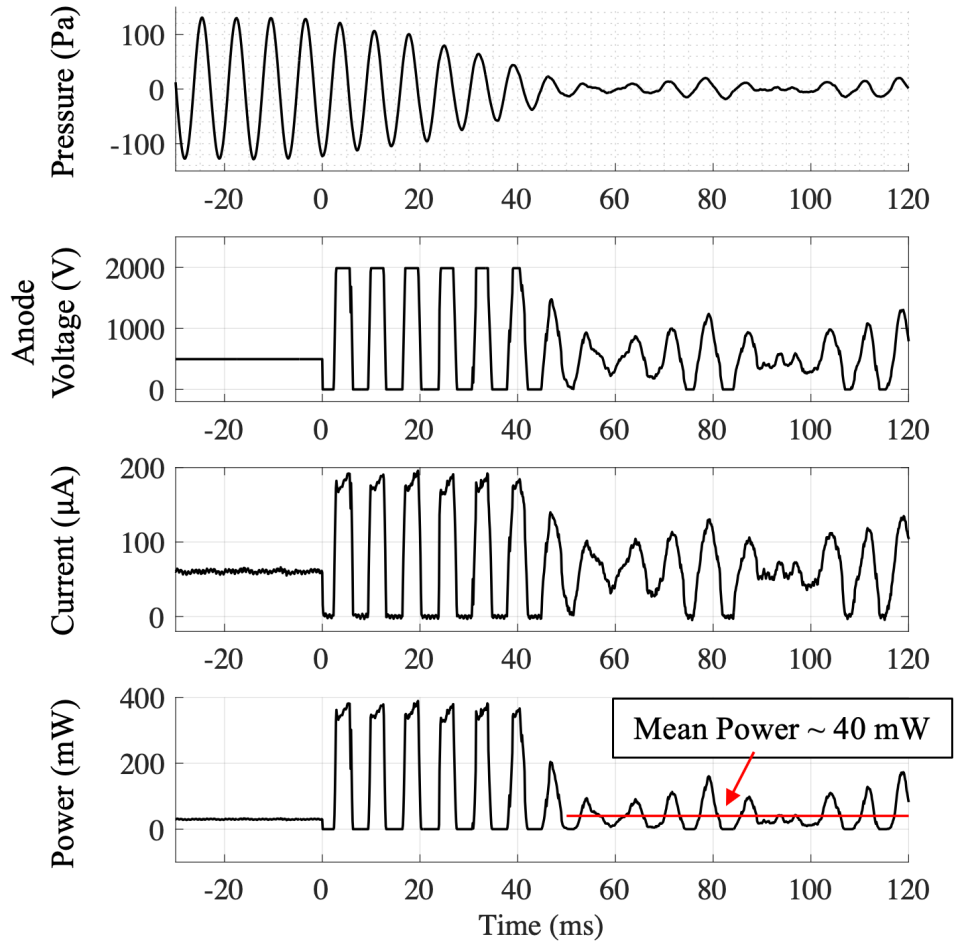

**Fig. S2. Electrical power required to stabilize the flame.** The plots show the pressure, anode voltage, current through the flame, and calculated electrical power while suppressing the instability. Once suppressed, the average electrical power use is only 40 mW, which is very little compared to the 3.4 kW flame thermal power.